\input harvmac

\def \NS{{\rm NS}}

\def \ep{\epsilon}

\def \x {\xi}

\def \del {\partial}
\def \bd {\bar \partial }

\def \ha{{\textstyle{1\over 2}}}

\def \a {\alpha}

\def \chi {\chi}

\def \p {\phi}
\def \m {\mu}
\def \n {\nu}

\def \H {{\cal H}}
\def \R {{\cal R}}\def \bx {\bar \x}
\def \td {\tilde }
\def \d {\delta}

\def \inv {^{-1}}
\def \ov {\over }

\def \d {\delta}

\font\mybb=msbm10 at 12pt
\def\bb#1{\hbox{\mybb#1}}

\def\bR {\bb{R}}


\def \lr { \lref}

\def \pl {{ Phys. Lett. }}

\def \pr {{ Phys. Rev. }}

\baselineskip8pt
\Title{
\vbox
{\baselineskip 6pt{\hbox{}}{\hbox
{
OHSTPY-HEP-T-99-025}}{\hbox{}} {\hbox{
}}} }
{\vbox{\centerline { 
}
\vskip4pt
\centerline{ Curved branes from String Dualities}
\vskip4pt
\centerline { }}}\vskip -20 true pt
\centerline { 
G. Papadopoulos$^a$, J.G. Russo$^b$ 
and A.A. Tseytlin$^{c,}$\footnote {$^*$} 
{Also at Lebedev 
Institute, Moscow and Imperial College, London.} }

\medskip \bigskip

\centerline{\it ${}^a$ Department of Mathematics, King's College London,}
\centerline {\it Strand, London WC2R 2LS}
\centerline{\rm gpapas@mth.kcl.ac.uk 
}

\medskip

\centerline{\it ${}^b$ Departamento de F\'\i sica, Universidad de Buenos Aires,}
\centerline{\it Ciudad Universitaria, Buenos Aires}
\centerline{\rm russo@df.uba.ar  }
\medskip

\centerline{\it ${}^c$ Department of Physics, The Ohio State University,} 
\centerline{\it 174 West 18th Avenue, Columbus, OH 43210, USA}
\centerline{\rm
tseytlin@mps.ohio-state.edu }

\bigskip\bigskip
\centerline {\bf Abstract}
\medskip
\baselineskip12pt
\noindent

We describe a simple method for generating new string
solutions for which the brane worldvolume is a curved space.
As a starting point we use  solutions with NS-NS charges 
combined with 2-d CFT's representing different parts
of space-time. We illustrate our method with many examples, 
some of which are associated with conformally invariant 
sigma models. Using U-duality, we also obtain supergravity
solutions with RR charges which can be interpreted as 
D-branes with non-trivial worldvolume geometry. 
In particular, we discuss the case of a D5-brane wrapped
on $AdS_3 \times  S^3$,  a solution interpolating between 
$AdS_3 \times S^3 \times \bR^5$ and $AdS_3 \times
 S^3 \times  S^3 \times  \bR$, 
and a D3-brane wrapped over $S^3 \times  \bR$
  or  $AdS_2 \times  S^2$.
Another class of solutions we discuss involves NS5-branes 
intersecting over a 3-space and NS5-branes intersecting 
over a line. These solutions are similar to D7-brane
or cosmic string  backgrounds.

\Date {November 1999}

\noblackbox
\baselineskip 14pt plus 2pt minus 2pt
\lr\GR{G.W.~Gibbons, M.B.~Green and M.J.~Perry,
``Instantons and Seven-Branes in Type IIB Superstring Theory,''
Phys.\ Lett.\ {\bf B370} (1996) 37;
hep-th/9511080.
}
\lr \CT{M. Cveti\v c and A.A. Tseytlin, ``General class of BPS saturated 
dyonic black holes as exact superstring
solutions,'' 
\pl { B366} (1996) 95; hep-th/9510097;``Solitonic strings 
and BPS saturated dyonic black holes,''
\pr D53 (1996) 5619; hep-th/9512031.
}
\lr \call { C.G.~Callan, J.A.~Harvey and A.~Strominger,
``World sheet approach to heterotic instantons and solitons,''
Nucl.\ Phys.\ {\bf B359} (1991) 611.
}
\lr\duffu {M.J.~Duff, H.~Lu and C.N.~Pope,
``AdS(5) x S(5) untwisted'',
Nucl.\ Phys.\ {\bf B532} (1998) 181;
hep-th/9803061.
}

\lr\hpone{P.S. Howe and G. Papadopoulos, ``Ultraviolet behaviour
of two-dimensional nonlinear sigma models,'' Nucl. Phys. {\bf B289} (1987)
264; ``Further remarks on the geometry of two-dimensional nonlinear
sigma models,'' Class. Quantum Grav. {\bf 5} (1988) 1677.}

\lr\hptwo{P.S. Howe and G. Papadopoulos, ``Finiteness and anomalies in 
(4,0)-supersymmetric sigma models,'' Nucl. Phys. {\bf B381} (1992)
360; hep-th/9203070.}

\lr \fund{ A.~Dabholkar, G.~Gibbons, J.A.~Harvey and F.~Ruiz Ruiz,
``Superstrings and Solitons,''
Nucl.\ Phys.\ {\bf B340} (1990) 33.}

\lr\ptone{ G. Papadopoulos and A. Teschendorff, 
``Multiangle five-brane intersections,'' Phys. Lett. 
{\bf 443} (1998) 159; hep-th/9806191.}

\lr\pttwo{ G. Papadopoulos and A. Teschendorff, 
``Grassmanians calibrations and five-brane intersections;'' hep-th/9811034.}

\lr\greene{ B. Greene, A. Shapere, C. Vafa and S.-T. Yau,
``Stringy cosmic strings and non-compact Calabi-Yau manifolds",
Nucl. Phys. {\bf B337} (1990) 1.}

\lr\bps{ H.J. Boonstra, B. Peeters and K. Skenderis, ``Brane 
intersections, anti-de Sitter space-times and dual superconformal
theories,'' Nucl.Phys. {\bf B533}(1998) 127; hep-th/9803231.}

\lr\coto{P.M. Cowdall and P.K. Townsend, ``Gauged supergravity vacua
from intersecting branes,'' Phys.Lett. {\bf B429} (1998) 281; Erratum-ibid.
{ \bf B434}(1998) 458;
hep-th/9801165.}

\lr \bur{J.~de Boer, A.~Pasquinucci and K.~Skenderis,
``AdS/CFT dualities involving large 2d N = 4 superconformal
symmetry,''
hep-th/9904073.
}

\lr \hor { G.T.~Horowitz and A.A.~Tseytlin,
``On exact solutions and singularities in string theory,''
Phys.\ Rev.\ {\bf D50} (1994) 5204; hep-th/9406067.}

\lr\rutse{ J.G. Russo and A.A. Tseytlin, 
``Magnetic flux tube models in superstring theory",
Nucl. Phys. {\bf B461} (1996) 131; hep-th/9508068}

\lr \klim { C.~Klimcik and A.A.~Tseytlin,
``Exact four-dimensional string solutions and Toda like 
sigma models from 'null gauged' WZNW theories,''
Nucl.\ Phys.\ {\bf B424} (1994) 71; hep-th/9402120.}

\lr \nap { C.R.~Nappi and E.~Witten,
``A WZW model based on a nonsemisimple group,''
Phys.\ Rev.\ Lett.\ {\bf 71} (1993) 3751; hep-th/9310112.
}
\lr\papad{G. Papadopoulos, ``Rotating rotated branes", JHEP 9904:014 (1999); 
hep-th/9902166.}

\lr\vafwi{C. Vafa and E. Witten, ``A Strong Coupling Test of 
$S$-Duality", Nucl.Phys. {\bf B431} (1994) 3; hep-th/9408074.}

\lr\giveon{S. Elitzur, O. Feinerman, A. Giveon and D. Tsabar,
``String Theory on $AdS_3\times S^3\times S^3\times S^1$,'' 
Phys.Lett. {\bf B449} (1999) 180; 
hep-th/9811245.}

\lr \CT{
M.~Cveti\v c and A.~Tseytlin,
``General class of BPS saturated dyonic black holes as exact superstring solutions,''
Phys.\ Lett.\ {\bf B366} (1996) 95; 
hep-th/9510097.
``Solitonic strings and BPS saturated dyonic black holes,''
Phys.\ Rev.\ {\bf D53} (1996) 5619; 
hep-th/9512031.}

\lr\tts{A.A. Tseytlin,
``No-force condition and BPS combinations of p-branes in 11 and 10 
dimensions,''
Nucl.\ Phys.\ {\bf B487} (1997) 141; 
hep-th/9609212. }

\lr \yank{ N.~Itzhaki, A.A.~Tseytlin and S.~Yankielowicz,
``Supergravity solutions for branes localized within branes,''
Phys.\ Lett.\ {\bf B432} (1998) 298; 
hep-th/9803103.}

\lr\mina{
R.~Minasian and D.~Tsimpis,
``On the geometry of non-trivially embedded branes,''
hep-th/9911042.
}

\lr \papd{G. Papadopoulos and P.K. Townsend, 
``Intersecting M-Branes,'' Phys.\ Lett.\ {\bf B380} (1996) 273; hep-th/9603087.}

\lr \george{
G. Papadopoulos, 
``T-duality and the worldvolume solitons of
five-branes and KK monopoles,'' Phys.\ Lett.\ {\bf B434} (1998) 277; 
hep-th/9712162.}

\lr \mod { A.A.~Tseytlin,
``Extreme dyonic black holes in string theory,''
Mod.\ Phys.\ Lett.\ {\bf A11} (1996) 689; hep-th/9601177.
}

\lr \jan{B.~Janssen,
``Curved branes and cosmological (a,b)-models;''
hep-th/9910077.}

\lr \rus{J. Russo,
``Einstein spaces in five and seven dimensions and non-supersymmetric gauge theories,''
Phys.\ Lett.\ {\bf B435} (1998) 284; 
hep-th/9804209.}

\lr\brech{
D.~Brecher and M.J.~Perry,
``Ricci-flat branes;''
hep-th/9908018.}

\lr \fifa {J.M.~Figueroa-O'Farrill,
``More Ricci-flat branes;'' hep-th/9910086.}

\lr\mel{
V.D.~Ivashchuk and V.N.~Melnikov,
``Intersecting p-brane solutions in multidimensional gravity and M-theory;''
hep-th/9612089.
``Multidimensional sigma-models with composite electric p-branes,''
Grav.\ Cosmol.\ {\bf 4} (1998) 73; 
gr-qc/9705005.
M.A.~Grebeniuk, V.D.~Ivashchuk and V.N.~Melnikov,
``Multidimensional cosmology for intersecting p-branes with static internal spaces,''
Grav.\ Cosmol.\ {\bf 4} (1998) 145; 
gr-qc/9804042.}

\lr\gmt{J.P. Gauntlett, R.C. Meyers and P.K. Townsend, ``Supersymmetry 
of rotating branes,'' Phys.\ Rev.\ {\bf D59} (1999) 025001; hep-th/9809065.}

\lr\malda{J.M.~Maldacena, ``The large N limit of superconformal 
field theories and supergravity,'' Adv. Theor. Math. Phys. {\bf 2} (1998) 231;
hep-th/9711200.}

\lr \calm{ C.G.~Callan and J.M.~Maldacena,
``D-brane Approach to Black Hole Quantum Mechanics,''
Nucl.\ Phys.\ {\bf B472}, 591 (1996)
; hep-th/9602043.}

\lr\dlp{M.J. Duff, H. Lu and C.N. Pope, ``Supersymmetry without supersymmetry,''
Phys. Lett. {\bf B409} (1997) 136; hep-th/9704189.}

\lr \card { M.~Bourdeau and G.~Lopes Cardoso,
``Finite energy solutions in three-dimensional heterotic string theory,''
Nucl.\ Phys.\ {\bf B522} (1998) 137;
hep-th/9709174.}

\lr\tet{A.A. Tseytlin, ``On the structure of composite
black p-brane solutions 
and related black holes,'' 
Phys. Lett. {\bf B395} (1997) 24; 
hep-th/9611111} 

\lr \TTM{ 
A.A. Tseytlin, ``Extreme dyonic black holes in string theory,''
Mod. Phys. Lett. A11 (1996) 689, 
hep-th/9601177 }

\lr \pw{ 
D.~Amati and C.~Klimcik,
``Nonperturbative Computation Of The Weyl Anomaly For A
Class Of Nontrivial Backgrounds,''
Phys.\ Lett.\ {\bf B219} (1989) 443.
G.T.~Horowitz and A.R.~Steif,
``Space-Time Singularities In String Theory,''
Phys.\ Rev.\ Lett.\ {\bf 64} (1990) 260;
``Strings In Strong Gravitational Fields,''
Phys.\ Rev.\ {\bf D42} (1990) 1950.
A.A.~Tseytlin,
``String vacuum backgrounds with covariantly constant null
Killing vector and 2-d quantum gravity,''
Nucl.\ Phys.\ {\bf B390} (1993) 153;
hep-th/9209023.
}

\lr \witten { E.~Witten,
``Anti-de Sitter space and holography,''
Adv.\ Theor.\ Math.\ Phys.\ {\bf 2} (1998) 253;
hep-th/9802150.}

\lr \kkk { 
R.D.~Sorkin,
``Kaluza-Klein Monopole,''
Phys.\ Rev.\ Lett.\ {\bf 51}, 87 (1983).
D.J.~Gross and M.J.~Perry,
``Magnetic Monopoles In Kaluza-Klein Theories,''
Nucl.\ Phys.\ {\bf B226} (1983) 29. 
}

\lr \harm{A.A.~Tseytlin,
``Harmonic superpositions of M-branes,''
Nucl.\ Phys.\ {\bf B475} (1996) 149;
hep-th/9604035.}

\lr \khuri{ R.R.~Khuri,
``Remark on string solitons,''
Phys.\ Rev.\ {\bf D48} (1993) 2947.}

\lr \papa { G.~Papadopoulos and P.K.~Townsend,
``Intersecting M-branes,''
Phys.\ Lett.\ {\bf B380} (1996) 273;
hep-th/9603087.}

\lr \gau{ J.P.~Gauntlett, D.A.~Kastor and J.~Traschen,
``Overlapping Branes in M-Theory,''
Nucl.\ Phys.\ {\bf B478} (1996) 544;
hep-th/9604179.}

\lr \TT { A.A.~Tseytlin,
``Composite BPS configurations of p-branes in 10 and 11
dimensions,''
Class.\ Quant.\ Grav.\ {\bf 14} (1997) 2085;
hep-th/9702163.}

\lr \duf { M.J.~Duff, S.~Ferrara, R.R.~Khuri and J.~Rahmfeld,
``Supersymmetry and dual string solitons,''
Phys.\ Lett.\ {\bf B356} (1995) 479;
hep-th/9506057.}

\lr \gaup {J.P.~Gauntlett, G.W.~Gibbons, G.~Papadopoulos and
P.K.~Townsend,
``Hyper-Kaehler manifolds and multiply intersecting
branes,''
Nucl.\ Phys.\ {\bf B500} (1997) 133;
hep-th/9702202.}

\newsec{Introduction}

Branes whose worldvolumes are curved spaces have widespread applications
in string theory, M-theory and in
studies of four- and five-dimensional black holes.
In particular, they have been used to identify the non-perturbative
states of strings in lower dimensions in various compactifications. 
However, in most cases the branes wrapped over curved
spaces were treated
as probes and their back reaction 
on spacetime geometry was ignored.
This is a consistent procedure for
D-branes which are wrapped on various homology cycles
at small string coupling but in general this is not the case.
Another application of branes with curved worldvolume,
or \lq curved branes' for short, is in scenarios where the $3+1$
dimensional universe is regarded as a brane moving in a higher dimensional space-time.
In particular, in some models
one can introduce gravity localized on the brane and study
gravitational collapse. Cosmological models have also been investigated in this context.

Our aim is to provide a systematic method to construct such solutions.
For this we begin with configurations that have NS-NS charges only. The
advantage of starting with
NS-NS configurations is that the conformal invariance
of the associated sigma models can be used to check
the consistency
of the backgrounds in string theory.
In this way we construct
a large number of configurations that can be interpreted as curved
branes. We find that many of these curved brane configurations
are limits of intersecting brane configurations possibly
superposed with plane waves and KK-monopoles.
The curved brane configurations
can thus be constructed from the \lq elementary' solutions of the NS-NS sector
which are associated with conformally invariant sigma models.
These include the following:

(a) The plane wave \pw . 

(b) The fundamental string (T-dual to plane wave) 
\refs{\fund,\hor} .

(c) The NS5-brane \call\ associated with a generic
harmonic function.

(d) The KK-monopole \kkk\ 
(or `KK 5-brane' T-dual to the NS5-brane)
and many other hyper-K\"ahler metrics in four 
and eight dimensions.

The exactness of most of the above solutions in string theory
is based on the ultraviolet finiteness of (4,q)-supersymmetric
two-dimensional sigma models \refs{\hpone, \hptwo}.
There is an associated `smeared' solution
for many of the above configurations.
This mostly applies to plane wave, fundamental string, NS5-brane and KK-monopole
all preserving 1/2 of the maximal spacetime supersymmetry.
The `smearing' 
is achieved by choosing 
the harmonic functions for these solutions
to be invariant under some translational isometries, i.e.
to be 
harmonic in a {\it subspace}
of the transverse space of the original solution.
The smeared solutions
share the conformal invariant properties of the 
generic solutions.

The above `elementary' solutions can be directly 
superposed to construct other
solutions which preserve various fractions of spacetime
supersymmetry. All these solutions are associated to intersecting M-brane
configurations \refs{\papa, \harm}. This can be achieved in 
various ways, using, for example the
harmonic function rule of \harm\ or the calibration methods of 
\pttwo .
The simplest solutions of that type are 
branes localized in 
all common transverse dimensions, 
but, in general, 
one can localize the brane of a lower dimension in 
the one of a higher dimension 
(the lower-dimensional brane satisfies the 
Laplace equation in curved geometry of the 
higher-dimensional brane \refs{\mod}). 
Some of the combinations that are known to be 
conformal to all orders in $\a'$ are the following:

(a) The fundamental string superposed with a plane wave \hor.

(b) The fundamental string/NS5-brane system, 
the NS5-brane superposed with a plane wave \refs{\CT,\mod}
and various T-dual configurations which include
the fundamental string/KK-monopole and the KK-monopole superposed
with a plane wave.

(c) The direct sum of two NS5-branes \khuri\ and the
intersection of two NS5-branes on a line.

(d) Intersecting NS5-branes at $Sp(2)$-angles on a line and
toric hyper-K\"ahler metrics in eight dimensions \refs{\gaup, \ptone}.

These are mostly the solutions that preserve $1/4$ of the spacetime supersymmetries,
apart from the last case which consists of solutions that preserve
at least a fraction $3/32$.

An alternative way to constructing solutions,
which leads to a broader class of solutions, 
can be obtained by ``curving" the transverse space of a plane wave 
or a fundamental string by putting there any CFT with 
correct central charge and solving the corresponding 
curved space Laplace equations for harmonic functions.
This general point of view and examples are 
already discussed in \refs{\hor,\mod}.
In the case of the NS5-brane, one 
can also curve the six worldvolume directions 
of the NS 5-brane or those of the KK monopole 
by putting there any CFT 
with the required central charge. For example,
this can be achieved using
a WZW model (possibly with
a linear dilaton to preserve the value 6 of the 
central charge of the internal space), or a plane wave or a 
fundamental string localized on the NS5-brane,
or, to leading order in $\a'$, 
any Ricci-flat space of dimension 1+5. 

{}From the above solutions of the NS-NS sector
one can easily find, using S- and T-dualities, many others
which have RR charges. This is in parallel to the construction
of the D5+D1 
solution which was found \calm\ by 
applying S-duality to the previously known 
$NS5+NS1$ configuration \mod. In particular, this 
will lead to new curved D-brane solutions in ten dimensions.
The eleven-dimensional counterpart of such 
solutions will be generalizations of M2 and M5
solutions having curved internal and/or external parts. 

To summarize, the key advantage of the approach based on
starting with an
NS-NS solution 
is that one does not need to solve the 
supergravity equations directly.
In fact, one can use simple 
CFT composition rules in NS-NS sector 
and then apply T- and S- dualities. 
In addition, the supersymmetry of some of the resulting 
solutions is easy to show in 
the sigma model framework.

Applying this method, 
we will find new solutions that include D-branes wrapped over a 
group space
or a coset space. For example, we will show that
there is a curved D5-brane solution
with worldvolume $AdS_3 \times S^3$
(with an extra RR 2-form background).
Another example is a curved
D3-brane with the world-volume being 
the
Nappi-Witten \nap\ space
or the 4-d null coset of \klim. In addition, we will
find a 
Euclidean curved D3-brane with worldvolume 
$R \times S^3$ and with linear dilaton. 

We shall also consider smeared NS5-branes, 
NS5-brane intersections on 3-dimensional space or on a line and 
a similar intersection with one NS5-brane replaced by a KK-monopole. 
Many of these solutions
are determined by harmonic functions on a two-dimensional transverse space.
Apart from the smeared NS5-branes, such solutions and their U-duals are 
solutions to the
leading-order in $\alpha'$ and they can be obtained from 
known M5-brane intersecting solutions
\refs{\papa,\harm,\gau}
by dimensionally reducing to type IIA D=10 theory along 
an overall transverse direction.\foot{The
multiple NS5-brane configuration intersecting 
over a 3-dimensional space can also be obtained, by 
smearing and U-duality, from 
a combination of a fundamental string and a plane 
wave \tet. However,
although the latter defines a conformal sigma model, the U-dual
configurations may receive higher order 
$\a'$-corrections.}
Such solutions will be shown to admit
a certain complex structure, and they are similar to 
the cosmic string solution of 
\greene. Consequently, their overall transverse directions can be
curved using the freedom of 
choosing the harmonic functions in two dimensions.
The D7+D3 bound state is U-dual to the intersecting NS5-brane configuration
on a 3-dimensional space. A question that remains open is whether the 
D7+D3 bound state is an exact string solution.

\newsec{Brane solutions with curved internal and transverse space}
\subsec{The ansatz}
{}From the NS-NS sector branes, 
the fundamental string and plane wave are the natural 
starting points for constructing solutions with curved
transverse space, while the NS5-brane and the KK monopole are
more appropriate for finding brane solutions with 
curved worldvolume.
To describe the configuration that describes both 
possibilities in the NS-NS sector,
we consider an eight-dimensional manifold $M$
with metric $d\hat s^2$, 
dilaton $\hat \phi$ and closed three-form $\hat H$. 
Then we write the ansatz \mod
\eqn\mone{
ds^2=2 g_1^{-1} du (dv+A+g_2 du)+d\hat s^2\ ,
}
$$
H=du\wedge dv\wedge dg_1^{-1} + 
du\wedge d(g_1^{-1} A)+ \hat H\ ,
$$
$$
e^{2\phi}= g_1^{-1} e^{2\hat \phi}\ ,
$$
where $g_1$ and $g_2$ are functions on $M$ associated with 
the fundamental string 
and the plane wave, respectively, $A$ is a (locally)
defined one-form on $M$ associated
with a rotation of the string. In this ansatz, the NS5-brane solution and possible
NS5-brane intersections or any other NS-NS background that allows a fundamental
string superposition are described by the geometry of $M$.
To find the string backgrounds described by this ansatz, we either
substitute \mone\ into the supergravity field equations or use 
the conformal invariance of sigma models to investigate the consistency
of the propagation of a string probe. The former coincides with the leading
order approximation in $\alpha'$ of the latter. The supergravity Killing
spinor equations for \mone\ were investigated in \papad\ and many new
supersymmetric solutions
were given.

The dynamics of a string probe in the NS-NS background \mone\ 
is described by a two-dimensional
sigma model with Lagrangian
\eqn\lag{
L_{10}= g_1^{-1} [\partial u \bar\partial v+ A_a
\partial u \bar\partial x^a+g_2 \del u \bar \del u ]+\gamma_{ab}
\partial x^a \bar\partial x^b}
$$
+\ \hat B_{ab} \partial x^a \bar\partial x^b
+ \R (-{1\over2} {\rm log}\ g_1+ \hat \phi)\ ,
$$
where $\cal R$=$ \a' R^{(2)}$.
The vanishing of the beta-functions in the leading order in $\alpha'$
gives
\eqn\mtwo{
\partial_a(\sqrt{\gamma} e^{-2\hat\phi} 
\gamma^{ab}\partial_b g_1)=0\ ,
\ \ \ \ \ \ 
\partial_a(\sqrt{\gamma} e^{-2\hat\phi} \gamma^{ab}\partial_b g_2)=0\ ,
}
$$
\partial_b (\sqrt{\gamma}e^{-2\hat\phi}F^{ba})
-{1\over2}\sqrt{\gamma} e^{-2\hat\phi}\hat H^{bca} F_{bc}=0\ ,
$$
$$
\hat R_{ab}-{1\over4}\hat H_{ace} \hat H_b{}^{ce}+2\nabla_a\partial_b \hat\phi=0\ , 
\qquad \partial_a
(\sqrt{\gamma} e^{-2\hat\phi}\hat H^{abc})=0\ ,
$$
where
$$
F_{ab}= 2\partial_{[a} A_{b]}\ ,\ \ \ \ \ \ 
d\hat s^2=\gamma_{ab} dx^a dx^b\ .
$$
These are precisely the field equations of the common sector of
supergravity theories adapted to the ansatz. 

\subsec{Conformal curved branes }

Some of the solutions described by the ansatz above
are exact to all orders in $\alpha'$.
To give some examples, we begin with the NS5-brane background for which
the associated sigma model is \call\ 
($\m=0,1,...,5;
\ m,n=1,...,4$)
\eqn\fiif{
L_{10} = L^{(0)}_6 (x) + L_4 (y) = 
\del x^\m \bd x^\m + (H_5 \d_{mn} + B_{mn})(y)
\del y^m \bd y^n + \R \p(y) \ , 
}
$$ e^{2\p} = H_5 (y) = 1 + {P \ov y^2} \ , \ \ \ \ \ \ \
dB = *d H_5 \ . 
$$
This is a conformal model because the
interacting part $L_4$ admits (4,4)
two-dimensional supersymmetry \hpone.
The same applies to many parallel NS5-branes for which $H_5$ is
a generic harmonic function on $\bR^4$.

Next, we can replace the 6-dimensional flat Minkowski worldvolume
directions of the NS5-brane with any conformal sigma model
with the same central charge. It is clear that the new sigma model
will be again conformal. 
Some examples are:

(1) A plane wave propagating in the NS5-brane world-volume
($i=1,2,3,4$)
\eqn\lls{
L_6 = \del u \bd v + g_2(u,x) du^2 + A_i(u,x) \del u \bd x^i + 
\del x^i \bd x^i \ , \ \ \ 
}
where $g_2$ is a harmonic function, e.g., 
$g_2=K= 1 + {Q \ov x^2}$. With this choice of $g_2$ the plane 
wave is localized on the worldvolume of NS5-brane but it is {\sl not}
localized along its transverse directions.

(2) A WZW model with central charge $c=6$.
In particular, one can take the Lagrangian 
$L_6$ to be that of $SL(2,\bR ) \times SU(2)$ WZW model
which in supersymmetric case has the same $c$ as 
the free six dimensional theory. 
Other examples include the $\bR^{1,2} \times SU(2)$ WZW model with linear dilaton in 
a spatial direction or 
some (super)cosets.

(3) The Nappi-Witten \nap\ WZW model supplemented with a
free CFT of two scalars. 

(4) In general, $L_6$ is allowed to depend on the transverse coordinates
of NS5-brane. For example, the function $g_1$ in \mone\ associated with the
fundamental string is allowed to depend on worldvolume $x$ and
transverse $y$ coordinates \mod. In particular, we get 
\eqn\fif{
L_{10} = g_1^{-1} (x,y) \del u \bd v + \del x^i \bd x^i
+
(H_5 \d_{mn} + B_{mn})(y)
\del y^m \bd y^n + \R \p \ , 
}
$$ e^{2\p} = H_5 (y) g_1^{-1} (x,y) \ , \ \ \ \ 
$$
where $H_1=g_1$ satisfies 
\eqn\eqq{
[\del^2_y + H_5(y) \del^2_x] H_{1}(x,y) =0\ .
}
There are many solutions to this equation. One is that of
a string smeared
over the NS5-brane but localised in 
transverse directions,
$$
g_1=H_1= 1 + { Q\ov y^2}, 
$$
which is the standard case, 
and another is of a string localised on
NS5-brane but smeared in all of the transverse directions, 
$$ 
g_1=H_1 = 1 + { Q\ov x^2} \ . 
$$
This latter case is T-dual to 
`wave on the NS5-brane' example mentioned above.

(5) Another choice is a magnetic field on the brane,
based on the
CFT investigated in \rutse. It is a solvable 2-d 
theory defined by the Lagrangian ($i=0,1,2$) 
$$ 
L_6=\del \rho \bar \del \rho + F(\rho)\rho^2 \big[\del \varphi
+(b_1+b_2)\del y\big] \big[\bar \del \varphi
+(b_1-b_2)\bar \del y\big]
$$\eqn\mgn{
+\ \del y\bar\del y +\del x_i \bar\del x_i +
 {1\over 2} {\cal R} \log F
\ ,
}
where 
$$
F={ 1 \ov 1+b_2^2\rho^2 } \ . 
$$
Here $b_1,b_2$ are constants, 
$y$ is a periodic coordinate of radius $R$, and $\rho ,\varphi $ 
are polar coordinates. The model describes a magnetic flux tube
in a Melvin-type geometry. The curvature of the
geometry is caused
by the magnetic field energy density.
There are two 
magnetic fields,
corresponding to the $U(1)\times U(1)$ gauge fields
associated to the components $g_{y\varphi}$ and $B_{y\varphi }$.
The flux lines are confined to a radius $\rho\sim 1/b_2 $
and $\rho\sim 1/b_1$, respectively.
Although there is no residual supersymmetry, the sigma model
is conformal to all orders in $\a' $.

(6) A NS5-brane placed on the worldvolume of the original NS5-brane.
The most general solution of this type is the one 
constructed in \refs{\ptone, \pttwo}
with the interpretation of NS5-branes 
intersecting on a line at $Sp(2)$
angles. The corresponding 2-d sigma model
has at least (4,0) supersymmetry, 
depending
on the choice of rotational parameters, and 
therefore is 
finite
to all orders in $\alpha'$. 
However, it may still receive $\alpha'$
corrections produced by finite counterterms 
which one needs to add to ensure 
cancellation of sigma model and supersymmetry
anomalies \hptwo. In the special case of 
{\it real} rotational parameters, the
sigma model supersymmetry is (4,4) 
and the background does not receive
$\alpha'$ corrections.

The above list can be enlarged by solutions 
which solve the supergravity equations 
only to leading $\a' $ order. 
In particular, one can take
any Ricci-flat 1+5 dimensional 
space, i.e.
$$
L_6=g_{\mu\nu}(x) \del x^\mu \bar \del x^\nu\ ,\ \ \ \ \ 
R^g_{\mu\nu}=0\ .
$$
Solutions of that type were discussed in 
\refs{\mel,\brech,\jan,\fifa}. A particular
class of them contains various
four-dimensional hyper-K\"ahler metrics supplemented with the two-dimensional
Minkowski space. Such backgrounds are exact CFT's.

\subsec{Conformal branes with curved transverse space}

In order to find brane solutions with curved transverse space, it is convenient
to begin with the fundamental string solution
\eqn\qqk{ L_{10} = g_1^{-1} (x) \del u \bar \del v 
- \ha \R \ln g_1 
+ L^{(0)}_8 (x) \ , \ \ \ \ \ 
L^{(0)}_8 = \del y^a \bar \del y^a \ . }
Then we can substitute the flat transverse directions with some CFT.
For example, the $SU(2)$ WZW model 
with linear dilaton, 
or NS5-brane and others.
The explicit expression for $g_1$ is then found to the leading order
in $\alpha'$ by solving the field equations \mtwo\ which in this case
reduce to 
$$
\partial_a \big(\sqrt{\gamma} \gamma^{ab} e^{-2\hat\p} \partial_b\big) g_1 =0 \ . 
$$
Some explicit examples involving 
WZW-type backgrounds were discussed \hor.
To the 
leading order in $\a'$, the transverse part
can be replaced by any Ricci flat metric,
leading to the following background 
\eqn\rrra{
\eqalign{
ds^2&=g_1^{-1}(y) (-dt^2+dx^2)+ \gamma_{ab}(y)dy^ady^b\ ,
\cr
B_{tx}&= g_1^{-1}\ ,\ \ \ \ \ \ 
e^{-2\phi}=g_1\ ,
\cr
R^\gamma_{ab}&=0\ ,\ \ \ \ \ \del_a (\gamma^{ab}\sqrt{\gamma}\del_b) g_1(y) =0\ .}
}
Another class of solutions can be obtained 
to the leading order in $\alpha'$
by taking the transverse space to be the configuration of 
NS5-branes intersecting 
at $Sp(2)$
angles on a line \ptone. To the same order, such solutions
can be superposed with a plane wave and rotation giving an even larger
class of backgrounds \papad. 

\subsec{Brane solutions generated by U-dualities}

Using T- and S-dualities we can construct configurations with
both NS-NS and R-R charges from the solutions
of the previous sections which have only NS-NS charges.
This method is well known and we shall not present
an exhaustive catalogue of all possibilities. Instead, 
we shall present a few examples.
Applying S-duality transformation to the model \fiif\ with 
the longitudinal part 
given by any of the models $L_6$ in section (2.1),
one obtains a curved D5-brane 
solution.
For example, if we choose the wave model \lls,
we find a solution describing a 
D5-brane with a wave localized on it. 
Other similar D-brane solutions can be constructed by
applying T-duality.
For example, if we assume that $g_2=K$ 
depends only on $u$ and $x^a=(x^1,x^2)$ but not on 
$z^k=(x^3,x^4)$ (i.e. it describes a wave 
in four dimensions smeared in the other two), 
thgn T-duality along $x^3,x^4$ gives a D3-brane 
solution
with a wave localized on it (with the whole configuration smeared 
in the two $z^k$ directions). The metric is 
given by
$$
ds^2_{10} = H^{-1/2}_5 (y) [ dudv + K(u,x)du^2 + dx^a dx^a]
+ H^{1/2}_5 (y)[dz^k dz^k + dy^n dy^n] \ . 
$$
Similarly, S-duality applied to the solution
\fif\ gives a D5 brane with D1 localized on it, i.e. 
$$
ds^2_{10} = H^{-1/2}_5 (y) [ H_1^{-1}(x) dudv + dx^s dx^s]
+ H^{1/2}_5 (y) dy^n dy^n\ .
$$
A $p_1$ brane within a $p$-brane, with $p_1<p$, can also be 
constructed by U-dualities. The solutions
for these 
composite branes have the 
following generic form 
$$ 
ds^2=H^{-1/2}_p(r) \big[ H_{p_1}^{-1/2}(\td r)(-dt^2+dx_1^2+...+dx_{p_1}^2)+ 
H_{p_1}^{1/2}(\td r) \big( d\td r^2+ \td r^2 d\Omega_{p-p_1-1}^2 \big)\big] 
$$ 
\eqn\ramo{ 
+\ H_{p_1}^{1/2}(\td r) H_p^{1/2}(r) \big( dr^2+ r^2 d\Omega_{8-p}^2 \big)\ , 
} 
\eqn\dig{ 
e^{-2(\phi -\phi_\infty) }=H_p^{(p-3)/2}(r ) H_{p_1}^{(p_1-3)/2}(\td r )\ \ , \ \ \ \ 
H_p(r)=1+{c_p \over r^{7-p} }\ ,
}
\eqn\ggh{
F_{8-p}=Q \ \Omega_{8-p}\ ,\ 
\ \ \ F_{p-p_1-1}= *\ \td \Omega_{p-p_1-1}\ . 
}
This solution is valid for $p-p_1=4$, $p_1<3$.

One can also construct curved Dp-brane solutions of the form
\eqn\uno{
ds^2= H_p^{-1/2}(y) g_{\mu\nu}(x) dx^\mu dx^\nu +
H_p^{1/2}(y) h_{mn}(y) dy^m dy^n
}
where $H_p(y)$ is an harmonic function in the curved space 
$ds^2_{9-p}=h_{mn}(y) dy^m dy^n$, which is transverse to the brane,
and 
$g_{\mu\nu}(x), h_{mn}(y) $ are Ricci-flat metrics, 
\eqn\dos{
R_{\mu\nu}^g(x)=0\ ,\ \ \ \ \ R_{mn}^h(y)=0\ .
}
Analogous solutions can 
be constructed in eleven-dimensional supergravity.
In particular, the following 
background represents a curved M5-brane: 
\eqn\mmff{
ds^2_{11}= H^{-1/3}(r)\ 
g_{\mu\nu}(x)dx^\mu dx^\nu +H^{2/3}(r) \big( dr^2 
+r^2 d\Omega^2_4 \big)\ , 
} 
$$ 
dC_3=6Q\Omega_4\ ,\ \ \ \ H(r)=1+{Q\ov r^3}\ ,\ \ \ \ 
R^g_{\mu\nu}=0 \ . 
$$
When the metric \mmff\ has translational isometry, this background can be connected to the previous backgrounds upon dimensional reduction and suitable U-duality transformations. 

We would like to emphasize, however, that 
the cases with Ricci flat longitudinal and/or transverse
spaces 
are just a small subclass of more general 
backgrounds. These more general configurations 
describe various interesting cases which include 
branes on group spaces, brane 
black holes and 
cosmological models with non-constant dilaton.

\newsec{ Some particular cases}

\subsec{D5-brane wrapped on $AdS_3\times S^3$ }

One example is D5-brane wrapped on $AdS_3\times S_3$. This
solution can be
obtained by an S-duality transformation applied to the NS5-brane model 
described in the previous section,
with worldvolume $L_6$ given by the $SL(2,\bR ) \times SU(2)$ WZW model.

An alternative derivation, which is also useful for further generalizations,
can be given by starting with the system
$ 1_{\NS}+ 5_{\NS} +5'_{\NS}$, with metric, 2-form and dilaton given by \refs{\TT}
\eqn\wwa{
ds^2=g^{-1}_1 (x,y) \big( -dt^2+ dz^2\big) +
H_5 (x) d x^n d x^n +
H'_5 (y) d y^m dy^m \ ,
}
\eqn\bbb{
dB = dg^{-1}_1 \wedge dt \wedge dz + *d H_5 + *dH'_5 \ , 
}
$$
e^{2\phi}={H_5(x) H_5(y)\over g_1(x,y)}\ ,
$$
where
\eqn\eeq{
[ H_5'(y)\del^2_x + H_5(x)\del^2_y]g_1(x,y)=0 \ . }
A particular solution is
\eqn\uiu{
g_1 (x,y) = H_1(x) H_1'(y) \ 
}
with
\eqn\oppp{
H_1=1+{R_1^2\over x^2}\ ,\ \ \ 
H_5=1+{R_5^2\over x^2}\ ,\ \ \ 
H'_1=1+{{R'_1}^2\over y^2}\ ,\ \ \ 
H'_5=1+{{R'_5}^2\over y^2}\ .
}
which corresponds to the ``dyonic string" 
generalization \duf\ of the $5_{NS}+5_{NS}$ solution 
of \khuri.
Different choices of the harmonic functions give
a number of interesting models; related solutions 
have been discussed in \refs{\gaup, \coto,\bps , \gmt ,\papad }.

To continue, let us further choose $H'_1=1$, in which case the solution
is just a direct product of 
NS1+NS5 configuration and another NS5-brane. 
Then we use the fact that the near-horizon limit of the NS1+NS5 configuration 
is simply the $SL(2) \times SU(2)$ WZW theory \mod.
Explicitly, as
$x\to 0$, the metric \wwa\ becomes 
\eqn\nnhh{
ds^2= {x^2\over R_1^2}(-dt^2+dz^2)+R_1^2{dx^2\over x^2}+
R_1^2d\Omega_3^2+ H'_5 (y) 
\big( dy^2+ y^2 d {\Omega'_3}^2\big) \ , 
}
where we have set for simplicity $H_1=H_5$.
The resulting 
sigma model describes a curved NS5-brane with worldvolume $AdS_3 \times S^3$
and defines an exact CFT. 
Applying S-duality, we obtain the type IIB solution
representing a curved D5-brane with worldvolume
$AdS_3\times S^3$ space. The associated metric is 
\eqn\nndd{
ds^2 = {H'_5}^{-1/2}\bigg[ 
{x^2\over R_1^2}(-dt^2+dz^2)+R_1^2{dx^2\over x^2}
+ R_1^2 d \Omega_3^2 \bigg]
+ {H'_5}^{1/2} 
\big( d y^2+ y^2 d {\Omega'_3}^2\big)\ . 
}
which is supplemented with the obvious 
dilaton and RR 2-form backgrounds.

The above solution may have applications to 
the study of
the decoupling limit of branes wrapped on curved spaces.
Other solutions can be constructed by applying T-duality along
isometric directions of $AdS_3 \times S^3$.

\subsec{D1-D5-D5 system with 
$AdS_3\times S^3\times S^3\times \bR $ horizon geometry}
Construction of another example requires to begin again 
with \wwa,\uiu\ but now choose 
$H_1=H_5$ and $H'_1= H'_5$ in which case 
$R_1=R_5\equiv R$ and $R'_1= R'_5\equiv R'$, see also \coto.
Now we take the near-horizon limit $y\to 0$.
Rescaling the coordinates, 
the metric becomes
\eqn\ddff{
ds^2=H_1^{-1} {u^2\over R'^2}
\big( -dt^2+dz^2\big) + R'^2 {d u^2\over u^2}
+H_1 \big( dx^2 +x^2 d\Omega_3^2 \big)
+ {R'}^2d {\Omega'_3}^2\ .
}
Since the dilaton is constant, the S-dual metric describing
a D1-D5-D5 system is the same as \ddff . 
The corresponding RR 2-form field strength is given by \bbb.

The solution \ddff\ interpolates between the 
two exact conformal models, 
as it is easy to show. Indeed,
as $x\to \infty $, the metric approaches that of 
$AdS_3\times S^3\times \bR ^4$.
For $x\to 0$, the 
metric takes the form 
\eqn\adss{
ds^2={u^2{u'}^2\over R^2{R'}^2}
(-dt^2+dz^2)+ R^2{du^2\over u^2}+{R'}^2
{d{u'}^2\over {u'}^2}
+ R^2 d\Omega_3^2 + {R'}^2d{\Omega'_3}^2\ ,
}
where we have rescaled the coordinates.
Introducing new variables $Y$ and $X$ by 
$$
u = R \exp \big[ {R_0\over R^2}Y+{ R_0 \over R R' } X \big]\ ,\ \ 
\ \ u' = R' \exp \big[ 
{ R_0\over {R'}^2}Y - {R_0\over R R' } X \big]\ ,\ \ \ 
{ 1 \ov R^2_0} ={ 1 \ov R^2} + {1 \ov {R'}^2} \ ,
$$
we get
\eqn\adss{
ds^2= e^{2Y\over R_0}
(-dt^2+dz^2)+ dY^2 + dX^2 
+ R^2 d\Omega_3^2 + {R'}^2d{\Omega'_3}^2\ \ . 
}
This is a 
direct product of the 3-dimensional anti-de Sitter space
of radius $R_0$, the two 3-spheres with radii $R$ and $R'$, respectively,
and a real line (parametrized by $X$).
String theory on such space and its duality to 
2-d superconformal theories was investigated in \bur.
Similar models like that of
$AdS_3\times S^3\times S^3\times S^1$ were 
recently studied also in \giveon.



\subsec{D3 brane wrapped over non-trivial 4d spaces}

It is straightforward to find
the solution
representing a D3 brane
``wrapped'' over a Ricci-flat
space.
In the `decoupling limit of \malda , 
we obtain ($R^2=\sqrt{4\pi g_s N}$, $\a'=1$) 
\eqn\neee{ 
ds^2= {r^2\ov R^2}\ g_{\mu\nu}(x) 
dx^\mu dx^\nu + R^2 {dr^2\ov r^2}+ R^2 d\Omega_5^2 
\ , 
} 
with constant dilaton $e^{\phi}=g_s$ and $R^g_{\mu\nu}=0$. 
The metric is of the form $X_5\times S^5$, where $X_5$ 
is an Einstein space having a Ricci flat four-manifold 
with metric $g_{\mu\nu}$ as a boundary.
Type IIB superstring theory on this background 
\neee\ should be dual to a ${\cal N}=4$ $SU(N)$ 
Yang-Mills theory on a four-dimensional manifold 
described by $g_{\mu\nu }$ \witten.\foot{If 
one is interested in calculating 
the gauge theory partition function,
one can take the longitudinal space 
to be some hyper-Kahler manifold,
for which various properties of the 
instanton moduli space are known. In particular, for certain 
ALE spaces, or K3, exact information is available 
for any $SU(N)$ \vafwi . 
The partition function $Z$ could then be compared with the 
string theory partition function on this background.} 

Like the D5 brane, the D3 brane can be also 
wrapped over non-Ricci-flat spaces with non-trivial 
2-form field and/or dilaton.
One particular example is obtained
by taking the parallel 4-d space to be the 
Nappi-Witten (NW) space. The resulting metric is then 
$$
ds^2_{10B}=H^{-1/2}\big[dudv+dx_1^2+dx_2^2+2\cos u dx_1dx_2\big]
+H^{1/2} \big[ dr^2+r^2 d\Omega_5^2\big]\ .
$$
where $H=1+R^4/r^4$.
It represents a monochromatic gravitational wave,
which travels in a direction parallel to the D3 brane (though it is not localized on the brane).
A simple check of consistency 
of this solution is to smear it in two transverse 
directions and to apply T-duality along them.
This gives a D5-brane solution wrapped over NW $\times \bR^2$
space (which is obviously a solution being 
$S$ dual to direct product of NS 5-brane and the NW$\times \bR^2$
WZW model).

The D3-brane can also be wrapped over 
$AdS_3 \times \bR $ space with linear dilaton in the spatial
direction
(upon smearing in the two transverse directions and U-duality 
the corresponding NS model is the direct product of the 
NS 5-brane model with the 
$SL(2) \times \bR \times \bR^2$ WZW model with linear dilaton 
along an isometric spatial direction).
Similarly, the Euclidean D3 brane may be wrapped over 
$S^3 \times \bR $ space with a linear dilaton along
spatial direction to make total central charge of the 
``longitudinal" theory equal to 
its standard free-theory
value. The metric and dilaton are given by
\eqn\sph{
ds^2_{10B}=H^{-1/2}\big[ dz^2+R_0^2 d\Omega_3^2\big]
+H^{1/2}\big( dr^2+r^2d{\Omega_5'}^2\big)\ ,
}
\eqn\rdd{
\phi= R_0^{-1}\ z\ ,\ \ \ \ \ H=1+{R^4\over r^4}\ .
}

Let us now construct a D3 brane wrapped over 
$AdS_2 \times S^2$. This is related to the D5 brane wrapped on
$AdS_3\times S^3$ by T-duality along isometric directions of
$AdS_3\times S^3$, but it can also be constructed
directly from 
an exact conformal model as follows.
We start with the conformal $\sigma $-model representing
the intersection of an NS 5-brane, a
Kaluza-Klein monopole (with Kaluza-Klein coordinate $x^2$),
a fundamental string, and a wave along the string direction 
$x^1$ \CT . The NS and KK 5-branes lie on the directions 
$x^1, x^3,x^4,x^5,x^6$. T-duality in $x^2$ exchanges their charges.
In four dimensions, this background leads to the dyonic
BPS black hole (which contains the Reissner-N\" ordstrom 
solution as a particular case). Let us consider the case when all 
four charges are equal.
The near-horizon geometry is $AdS_2\times S^2\times \bR^6$, the 
$AdS_2\times S^2 $ part of the space being described by
the coordinates $x^0,x^7,x^8,x^9$.

In order to incorporate a D3 brane wrapped on $AdS_2\times S^2$,
we start with this conformal sigma model in the horizon limit
$AdS_2\times S^2\times \bR^6$.
We add an NS 5-brane lying along the space 
$(x^1,x^2,x^7,x^8,x^9)$. The transverse part of this extra
brane
is thus the flat space $(x^3,x^4,x^5,x^6)$, so that the
resulting metric is given by
$$
ds^2_{10B}= {1\over x^2_7}(-dx_0^2+R^2 dx_7^2) +
R^2d\Omega_2^2 +dx_1^2+dx_2^2+ 
H(r)\big[ dr^2+ r^2 d{\Omega'_3}^2\big]\ ,
$$
$$
e^{2\phi}=H=1+{Q\over r^2}\ ,\ \ \ \ r^2=x_3^2+x_4^2+x_5^2+x_6^2\ .
$$
Then we perform
S-duality transformation, and T-duality 
transformations in the directions $x^1$, $x^2$, 
which transform the extra NS5 brane into a D3 brane smeared
in the directions $(x^1,x^2)$, and the original brane
configuration
of \CT\ into a (near horizon) configuration representing
a D5 brane (lying on $x^2,x^3,x^4,x^5,x^6$~)
and an NS5 brane (on $x^1,x^3,x^4,x^5,x^6$~), 
a D1 brane (along $x^2$) and a fundamental string (along $x^1$). The 
resulting background is given by
$$
ds^2_{10B}=H^{-1/2}(r)\big[ {1\over x^2_7}(-dx_0^2+R^2 dx_7^2) +
R^2d\Omega_2^2 \big] $$
\eqn\ddad{
+\ H^{1/2}(r)
\big[ dx_1^2+dx_2^2+dr^2+ r^2 d{\Omega'_3}^2\big]\ ,
}
\eqn\mas{
H=1+{Q\over r^2}\ ,\ \ \ \ 
F_5=Q (1+*) (\Omega'_3\wedge dx_1\wedge dx_2)\ , \ \ \ \ 
e^{\phi}=g_s\ , }
\eqn\ggf{
H_3^{NS}= {1\over R} dx_0\wedge dx_1\wedge dx_7
+R\ \Omega_2\wedge dx_7\ ,
\ \ \ \ 
H_3^{R}= {1\over R} dx_0\wedge dx_2\wedge dx_7+R\
\Omega_2\wedge dx_7\ .
}

\subsec {D2-brane on $S^2$}

Many of the solutions that we have presented above contain the
metric of an odd-dimensional 
sphere $S^{2n+1}$. In all such cases, one can use the Hopf fibration 
$S^1\rightarrow S^{2n+1}\rightarrow CP^n$ 
and perform T-duality along $S^1$ in a way similar to that of \dlp. 
It turns out that for D-brane backgrounds in the dual picture
the Hopf fibration untwists and the sphere is replaced by
$CP^n$.

For an application, let us consider the solution \sph , \rdd .
This is a euclidean D3 brane wrapped on $S^3$. Next we use the
Hopf fibration $S^1\rightarrow S^{3}\rightarrow S^2$, $CP^1=S^2$,
to obtain a solution with the interpretation of a euclidean
D2 brane wrapped on $S^2$.
The metric and dilaton of this background is
\eqn\ttre{
ds^2_{10B}=H^{-1/2}\big[dz^2+R_0^2d\Omega_2^2\big]
+H^{1/2}\big(d\theta^2 + dr^2+r^2d\Omega_5^2\big)\ ,
}
$$
e^{2\phi }=e^{2z/R_0} H^{1/2}\ .
$$
The eleven-dimensional supergravity solution
representing an M2 brane wrapped on $S^2$ 
is then constructed 
by applying the usual formula of dimensional reduction.

\newsec{Solutions with 2-dimensional
transverse space }
Now we are going to consider a 
different 
class of NS-NS solutions which are determined by functions
which depend on two variables. Many examples of such solutions
have the interpretation of intersecting branes with a two-dimensional
overall transverse space.
We can either curve the worldvolume directions of the branes
involved in the intersection or the overall transverse space.
The investigation of the former case is similar to that presented
in the previous section and we shall not repeat the analysis.
In the latter case we first remark that 
any two-dimensional metric
is conformal to the flat one and that in two dimensions there is
a freedom of defining harmonic function by
adding the real part 
of a holomorphic one. A product of these harmonic functions
is the conformal coefficient of the overall transverse space.
As we shall see, 
making different
choices for the harmonic functions, we effectively curve the overall transverse
space of the configuration. This phenomenon is 
similar to what happens 
in the case of the cosmic string 
background \greene. In fact, we shall find that one
of our solutions reduces to that of the cosmic string. As a byproduct of
our analysis, we shall give two examples of Calabi-Yau metrics with holonomy
$SU(3)$ and $SU(4)$, respectively.

\subsec{Single ``smeared'' NS5-brane}

Let us start with a
NS5 brane solution smeared in {\it two }
transverse dimensions. As we have already mentioned, 
this
is an exact conformal model just like 
the standard
NS5 brane.
In particular, we smear the directions 
$\{\x_{a}; \ a=1,2\}=(x_3, x_4)$ and keep the dependence of the
solution on the coordinates 
$\{x_i; i=1,2\}$. In such case, the NS-NS 3-form field strength is 
\eqn\soo{
(dB)_{12i} = \ep_{ij} \del_j H \ , \ \ \
\ \ \ \ \del^i\del_i H=0 \ .
}
Next, we consider a holomorphic function $\H=H+i B$ of $z=x_1+ix_2$
that solves $\del \bd \H=0, \ \del = \del/\del z$. 
Then 
$$ \del_i H = \ep_{ij} \del_j B\ , $$ 
and so the non-zero $2 \times 2$ components
of $B_{mn}$ can be chosen as
\eqn\cho{
B_{ab} = \ep_{ab} B
\ . }
The resulting NS5 brane sigma-model can be written as
\eqn\modd{
L= -\del t \bd t + \del y_n \bd y_n
+ (H \d_{ab} + B\ep_{ab})(x) \del \x_a \bd \x_b
+ H(x) \del x_i \bd x_i + \ha \R \ln H (x)\ . 
}
Note that 
$$
(H \d_{ab} + B\ep_{ab})(x) \del \x_a \bd \x_b
\equiv 
\ha [\H (z) \del \x \bd \bx + c.c.]\ , 
$$ 
where 
$\x= \x_1 + i \x_2$. 
Since the 
torsion in two dimensions 
is trivial, i.e. adding $B_{ij}=\ep_{ij} B(x)$ 
gives a total 
derivative which we can drop, 
we can write this in a complex form
\eqn\hhha{
L= -\del t \bd t + \del y_n \bd y_n
+ \ha [\H (z) (\del \x \bd \bx 
+ \del z \bd \bar z) 
+ 
\bar \H (z) (\del \bar \x \bd \x 
+ \del \bar z \bd z) ] + \ha \R \ln H(x)
\ . 
} 
This background is related, via T-duality, to the D7 brane 
one \GR, 
and, via compactification -- to cosmic string
geometry of \greene.
More specifically, $T$-duality along 
a spatial 5-brane direction gives the IIB NS5 brane;
S-duality and T-duality along $\x_a$
gives the D7-brane. 
$\x_a$ are directions of 2-torus with trivial complex 
structure, and the 
background 
modulus field is $E_{ab} = H \d_{ab} + B\ep_{ab}, $
or $$\rho= \rho_1 + i \rho_2=
B + i (\det G_{ab})^{1/2 } = -i \H\ , \ 
\ \ \ \ \ \ e^{2\p} = H = \rho_2\ . $$
$T$-duality along one of the $\x_a$ directions gives 
the corresponding smeared version of the KK monopole:
$$
L= -\del t \bd t + \del y_n \bd y_n
+ H\inv (x) [\del \x_1 + B (x) \del \x_2][\bd \x_1 
+ B(x) \bd \x_2] $$
\eqn\ytr{
+ \ H(x) \del \x_2 \bd \x_2 + H(x) \del x_i \bd x_i 
\ , }
or 
\eqn\orr{
L= -\del t \bd t + \del y_n \bd y_n
+ H\inv (x) |\del \x_1 + \tau(x) \del \x_2|^2
+ H(x) \del x_i \bd x_i 
\ , }
where the modulus of the two-torus is $\tau = B + i H = \rho$. 
Compactifying to four dimensions, 
this reduces indeed
to the model of \greene, i.e.
$T^2 \times \bR^2$.

\subsec{NS5-brane and KK monopole}
One can also start with the 
conformal model describing
NS5 brane \ + \ KK monopole \ which is smeared 
in one transverse direction, say $y=x_3$. 
Then we find that 
$$
L= -\del t \bd t + \del y_n \bd y_n + 
H_1 (x) H\inv_2(x) \big[ \del z + a_3 (x) \del x_3 \big] \big[ 
\bd z + a_3 (x) \bd x_3 \big] 
$$ 
\eqn\one{
+\ H_1(x) H_2(x) (\del x_3 \bd x_3 + \del x_i \bd x_i) + 
b_3 (x) (\del z \bd x_3 - \bd z \del x_3)
+ {\cal R} \p(x) \ , 
} 
$$
\del_i\del_i H_1 = \del_i\del_i H_2 =0 \ , 
\ \ \ \ 
\del_{i} b_3 = \ep_{ij} \del_j H_1 \ , 
\ \ \
\del_{i} a_3 = \ep_{ij} \del_j H_2 \ , 
$$
where $i=1,2$ and we have made a natural 
special choice of the two-form gauge potential. 
We can further identify $b_3$ and $a_3$ with the imaginary 
parts of the holomorphic functions $\H_1$ and $\H_2$
that have $H_1$ and $H_2$ as their real parts.

This background is U-dual, in particular, 
to D6 + NS5 configuration with one isometric direction,
or 
to D7 + KK-monopole. 


\subsec{Intersecting NS5 branes}

Let us now consider the solution of two NS5 branes intersecting
on a 3-dimensional space and localized only in the overall two
transverse directions. The corresponding M-theory solution 
is that of
$M5 \bot M5$ \refs{\papa,\harm}
smeared in the 11-th
transverse 
direction. The NS5-brane system is related, by an S-duality and
a chain of two T-dualities, to a ``D3-brane on a D7-brane" configuration.
This is 1/4 BPS state and the solution is given 
by the harmonic function rule. The solution is smeared in all eight common
directions and depends only on the two overall transverse directions.
The sigma model action for this NS5-brane configuration is 
$$
L= -\del t \bd t + \del y_s \bd y_s + (H \d_{ab} + B \ep_{ab})(x)
\del \x_a \bd \x_b $$
\eqn\uyi{
+\ (H' \d_{ab} + B' \ep_{ab})(x)
\del \eta_a\bd \eta_b + H(x) H'(x) \del x_i \bd x_i + \R \p 
\ , } 
where $s=1,2,3, \ a,b=1,2, \ i=1,2, $ 
and the dilaton is 
$ e^{2\p} = H (x) H' (x)$.

In general, $H$ and $H'$ are two 
{\it arbitrary} real functions that solve 
the 2-d Laplace equation, while $B_{\m\n}$
is determined by 
\eqn\fre{
(dB)_{\x_1\x_2 i} = \ep_{ij}\del_i H \ , 
\ \ \ \ \ \ \ 
(dB')_{\eta_1\eta_2 i} = \ep_{ij}\del_i H' \ .
}
To put
the action in the above simple
local form we made the assumption
that $H$ and $H'$ are real parts of the
two holomorphic functions 
$\H,\H'$, and 
used the same considerations as above 
for a single NS5-brane 
smeared in two directions 
(which is an obvious special case $H'=1$) to
determine the antisymmetric tensor
in terms of $Im \H = B $ and $ Im \H'=B'$.
We remark that the 
$5_{\NS} \bot 5_{\NS}$ solution with a
different 
special choice
of $H$ and $H'$ (such 
that $H + i H'$ is a holomorphic function)
was considered in \george.

The resulting action has a remarkably simple structure 
$$
L= -\del t \bd t + \del y_n \bd y_n
+ \ha \bigg([\H (z) \del \x \bd \bx 
+ \H' (z) \del \eta \bd \bar \eta ] + c.c. \bigg)
$$
\eqn\rrty{
+\ \ha H(x) H'(x) (\del z \bd \bar z + c.c.)
+ \ha \R \ln [H(x) H'(x)] \ , 
}
where
$z=x_1 + i x_2, \ 
\eta=\eta_1 +i \eta_2, \ \x=\x_1 + i\x_2$
and $n=1,2,3$.\foot{Compactified to 3 dimensions 
this background 
gives a black hole solution with several scalars
(cf. \card).}
A special choice is $\H= Q \ln z, \ 
\H'= Q' \ln z$, so that $H = \ha Q \ln |z|^2, \ 
B = \ha i \ln (z/\bar z )$, 
and similarly for $B'$.

The corresponding type II supergravity background,
or the associated M-theory configuration,
preserves $1/4$ of 
the spacetime supersymmetry.
This can be seen using the holonomy
of the associated connections $\nabla^{(\pm)}$.
For this we introduce the complex structures
$
I(d\x)=i d\x ,\ 
I(d\eta)=i d\eta, \ 
I(dz)=i dz$
and
$
J(d\x)=i d\x ,\ 
J(d\eta)=i d\eta, \ 
J(dz)=-i dz.$
These complex structures are integrable
since they are constant.
A direct computation reveals that
$\nabla^{(+)} I=0$
and $\nabla^{(-)} J=0$ which implies that the holonomy
of $\nabla^{(\pm)}$ is a subgroup of $U(3)$. In fact, it turns
out that the holonomy of both connections is $SU(3)$. 
Using this one can conclude that
the solution indeed preserves $1/4$ of the
supersymmetry. As a consequence, the sigma model
\rrty\ admits a 
(2,2)-supersymmetric extension. 
This
amount of supersymmetry by itself is not, however, 
sufficient to prove 
that this sigma model is finite 
to all orders in perturbation theory since the 
argument of \hpone\ does not apply.
It may nevertheless be an exact CFT due to the simple
form of the action.

Another consequence of the $SU(3)$ holonomy is that these metrics are associated
with Calabi-Yau ones. In particular, let us T-dualize twice, once along
the $\x$ and once along the $\eta$ coordinates. Using, the results in section (4.1),
we find that 
$$
L= -\del t \bd t + \del y_n \bd y_n
+ H\inv (x) |\del \x_1 + \tau(x) \del \x_2|^2
$$
\eqn\orrr{
+ \ H'{}^{-1} (x) |\del \eta_1 + \tau'(x) \del \eta_2|^2
+ H(x)H'(x) \del x_i \bd x_i 
\ , 
} 
where $\tau=-i\H$ and $\tau'=-i\H'$. The non-trivial part of the
sigma model action is associated with a 
six-dimensional non-compact Calabi-Yau
metric.

\subsec{Three NS5 branes intersecting on a line}

The metric of the supergravity solution of three NS5 branes pairwise
intersecting on three-dimensional spaces, and all 
-- on a
line, is
\eqn\ppz{
\eqalign{
ds^2&= -dt^2 + dy^2 +H_1 d\xi d \bar \xi+
H_2 d\eta d \bar \eta +H_3 d\zeta d \bar \zeta 
+H_1 H_2 H_3 dz\bar dz\ . }
}
Here $\xi ,\eta ,\zeta$ and $z$ are complex
coordinates and $H_1,H_2, H_3$ are harmonic
functions in the overall transverse space spanned by $(z,\bar z)$.
This solution
may 
be viewed as a reduction of 
$M5\perp M5 \perp M5$ 
background in eleven dimensions \refs{\papa,\harm}. 
Performing similar calculations as in the previous cases, the
sigma model action is found to be 
$$
L= -\del t \bd t + \del y \bd y
+ \ha \bigg([\H_1 (z) \del \x \bd \bx 
+ \H_2 (z) \del \eta \bd \bar \eta +\H_3 (z) \del \zeta \bd \bar \zeta] + c.c. \bigg)
$$
\eqn\rrtyy{
+ \ \ha H_1(x) H_2(x) H_3(x) (\del z \bd \bar z + c.c.)
+ \ha \R \ln [H_1(x) H_2(x) H_3(x)] \ , 
}
where $\H_1, \H_2, \H_3$ are holomorphic functions of $z$. 
The above solution 
preserves $1/8$ of the
spacetime supersymmetry,
as can be seen by studying the holonomy 
of the connections $\nabla^{(\pm)}$. In particular, the 
holonomy of both connections
is $SU(4)$.
The corresponding 
bosonic sigma model admits a 
(2,2)-supersymmetric extension which again is not enough to
establish that the model is conformal using the argument of \hpone.

As in the previous case, the above NS5-brane geometry is associated to
an eight-dimensional Calabi-Yau
one. Indeed, using T-duality in three directions chosen in a way similar
to that of the previous example, we find
\eqn\orcrr{
\eqalign{
L&= -\del t \bd t + \del y \bd y
+ H_1\inv (x) |\del \x_1 + \tau_1(x) \del \x_2|^2
\cr &
+ 
H_2\inv (x) |\del \eta_1 + \tau_2(x) \del \eta_2|^2+
H_3\inv (x) |\del \zeta_1 + \tau_3(x) \del \zeta_2|^2
\cr &
+ H_1(x)H_2(x)H_3(x) \del x_i \bd x_i 
\ ,}
}
where $\tau_1=-i\H_1$, $\tau_2=-i\H_2$ and $\tau_3=-i\H_3$. The Calabi-Yau
metric is given by the non-trivial part of the sigma model action.

\newsec{Conclusions}
We have obtained 
brane solutions with a curved worldvolume
by starting with conformal 
 two-dimensional sigma models describing the
propagation of strings in NS-NS backgrounds.
The advantage of this method 
is the simplicity and transparency of the 
construction of brane solutions. 
Many of the solutions that have already appeared in the literature
\refs{\mel -\fifa} can then be easily derived by dualities,
 without solving any differential equations. 
One of the main points of the present
 discussion is that branes with a Ricci-flat worldvolume 
 constitute  only a  special case of a more general class of configurations.
The worldvolume of a brane can be any suitable CFT, which
may describe non-Ricci flat spaces with
non-constant dilaton and non-vanishing 2-form field strengths.
In particular, we have presented examples of 
the solutions  representing  branes
wrapped over group spaces.

Further progress in constructing 
interesting curved brane solutions  
 is related to the undestanding of 
the back reaction of branes 
 wrapped on  homology cycles
of  compact spaces in string and M-theory compactifications.
It is clear from our results that apart
from the topological information 
about the cycles one
 also needs some geometrical data. In some situations 
 the geometry of the normal bundle of the cycle will 
 also be relevant.\foot{In this paper we have 
assumed an orthogonal
decomposition between the worldvolume and
the  transverse spaces.
  In general,  
 if one wraps a brane  over 
a cycle  one  may  not  have an orthogonal
decomposition in the metric
 because the normal bundle of the cycle
can be twisted. Examples of 
such solutions were discussed in
\mina.}

\bigskip\bigskip
{\bf Acknowledgements}

\noindent
G.P. is supported by the Royal Society. J.R. is supported by UBA,
Conicet and Fundaci\' on Antorchas.
The work of A.T. is supported in part by
the DOE grant No. DOE/ER/01545-777, 
by the EC TMR programme ERBFMRX-CT96-0045, 
INTAS grant No.96-538,
and NATO grant PST.CLG 974965.

\listrefs
\end